# Automatic Data Deformation Analysis on Evolving Folksonomy Driven Environment

**Massimiliano Dal Mas**
me @ maxdalmas.com

**ABSTRACT**

The Folksodriven framework makes it possible for data scientists to define an ontology environment where searching for buried patterns that have some kind of predictive power to build predictive models more effectively. It accomplishes this through an abstractions that isolate parameters of the predictive modeling process searching for patterns and designing the feature set, too. To reflect the evolving knowledge, this paper considers ontologies based on folksonomies according to a new concept structure called "Folksodriven" to represent folksonomies. So, the studies on the transformational regulation of the Folksodriven tags are regarded to be important for adaptive folksonomies classifications in an evolving environment used by Intelligent Systems to represent the knowledge sharing. Folksodriven tags are used to categorize salient data points so they can be fed to a machine-learning system and "featurizing" the data.

*Keywords:* Computational Intelligence, Artificial Intelligence, Big Data, Sentiment analysis, Sentic computing, Semantic Web, Folksonomy, Ontology, Network, Elasticity, Plasticity, Natural Language Processing, Quasicrystal

**INTRODUCTION**

Computational Intelligence (CI) is a methodology to deal/learn with new situations for which there are no effective computational algorithms [1, 2, 3].

A fundamental prerequisite for an evolving environment is to decide the "knowledge" for a task domain: what kinds of things consists of, and how they are related to each other.

Semantic networks can be used to simulate the human-level intelligence providing efficient association and inference mechanisms to simulate the human complex frames in reasoning. Ontology can be used to fill the gap between human and CI for a task domain with an adaptive ontology matching.

The main purpose of this chapter is the development of a constitutive model emphasizing the use of ontology-driven processing to achieve a better understanding of the contextual role of concepts (Sentic computing). To reflect the evolving

---

knowledge this chapter considers an ontology-driven process based on folksonomies according to a new concept structure called "Folksodriven" [4 - 10] to represent folksonomies - a set of terms that a group of users tagged content without a controlled vocabulary. To solve the problems inherent an uncontrolled vocabulary of the folksonomy a *Folksodriven Structure Network* (*FSN*), built from the relations among the *Folksodriven tags* (*FD tags*), is presented as a folksonomy tags suggestions for the user. It was observed that the properties of the *FSN* depend mainly on the nature, distribution, size and the quality of the reinforcing *FD tags*. So, the studies on the transformational regulation of the *FD tag*s are regarded to be important for adaptive folksonomies classifications in an evolving environment used by Intelligent Systems.

This work proposes to use an adaptive ontology matching to understand what knowledge is required for a task domain in an evolving environment, using the elastodynamics – a mathematical study of a structure deformation that become internally stressed due to loading conditions – to obtain an *elasto–adaptative–dynamic methodology* of the *FSN*.

## DATA PREPARATION

Folksonomies are concentrations of user-generated categorization principles. An ontology is a formal specification of a conceptualization of an abstract representation of the world or domain we want to model for a certain purpose.

Ontologies can capture the semantics of a set of terms used by some communities: but meanings change over time, and are based on individual experiences, and logical axioms can only partially reflect them.In a model of space-time (Fig. 1), every point in space has four coordinates ($x, y, z, t$), three of which represent a point in space, and the fourth a precise moment in time. Intuitively, each point represents an event that happened at a particular place at a precise moment. The usage of the *four-vector* name assumes that its components refer to a "standard basis" on a Minkowski space [11]. Points in a Minkowski space are regarded as events in space-time. On a direction of time for the time vector we have:

- past directed time vector, whose first component is negative, to model the "history events" on the folksodriven notation
- future directed time vector, whose first component is positive, to model the "future events" on the folksodriven notation

$$(1) \quad FD := (C, E, R, X)$$

A Folksodriven will be considered as a tuple (1) defined by finite sets composed by the *Formal Context* (*C*), the *Time Exposition (E)*, the *Resource (R)* and the ternary relation *X* – as in Fig. 1.

As stated in [4 - 9] we consider a *Folksodriven tag* (*FD tag*) as a tuple (1) defined by finite sets composed by:

• *Formal Context (C)* is a triple $C:=(T, D, I)$ where the *Topic of Interest T* and the *Description D* are sets of data and *I* is a relation between *T* and *D*;

• *Time Exposition (E)* is the clickthrough rate (CTR) as the number of clicks on a

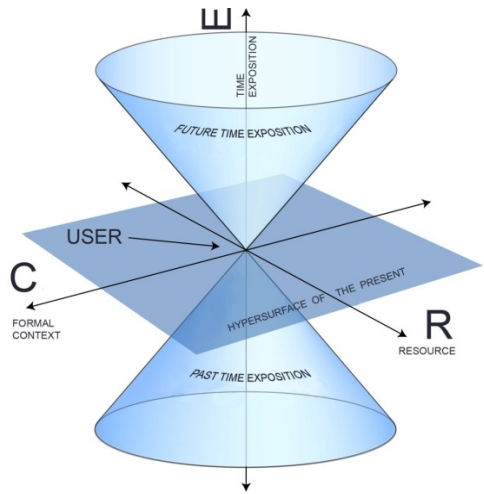

**Fig. 1.** In a model of space-time, every point in space has four coordinates ($C$, $E$, $R$, $X$), representing a point in space and a precise moment in time $E$

*Resource (R)* divided by the number of times that the *Resource (R)* is displayed (impressions);

• *Resource (R)* is represented by the URI of the webpage that the user wants to correlate to a chosen tag;

• *X* is defined by the relation $X = C \times E \times R$ in a Minkowski vector space [11] delimited by the vectors $C$, $E$ and $R$.

We consider a Folksodriven network in which nodes are Folksodriven tags (*FD tags*) and links are semantic acquaintance relationships between them according to the SUMO (http://www.ontologyportal.org) formal ontology that has been mapped to the WordNet lexicon (http://wordnet.princeton.edu). It is easy to see that *FD tags* tend to form groups, i.e. small groups in which tags are close related to each one, so we can think of such groups as a complete graph. In addition, the *FD tags* of a group also have a few acquaintance relationships to *FD tags* outside that group. Some *FD tags*, however, are so related to other tags (e.g.: workers, engineers) that are connected to a large number of groups. Those *FD tags* may be considered the hubs responsible for making such network a scale-free network [12].

The network structure of "Folksodriven tags" (FD tags) – Folksodriven Structure Network (FSN) – was thought as a "Folksonomy tags suggestions" for the user on a dataset built on chosen websites.

## FINDING PROBLEMS

With a Folksodriven Structure Network (FSN), time-series data can be represented in tables, where the columns contain measurements and the times at which they were made.

Traditionally, if we want to extract a diverse subset from a large data set, the first step is to create a similarity matrix — a huge table that maps every point in the data set against every other point. The intersection of the row representing one data item and the column representing another contains the points' similarity score on some standard measure.

There are several standard methods to extract diverse subsets, but they all involve operations performed on the matrix as a whole. With a data set with a million data points — and a million-by-million similarity matrix — this is prohibitively time consuming. The following method attempts to address the problem in a different way.

## DATA DEFORMATION ANALYSIS

An elastic lattice *FSN* exhibits deformation is connected to the relative movement between points in it that is represented by the displacement field [13]. Considering an elastic lattice *FSN1*, refer to Fig. 2, a *FD tag* defined in a cell *i* turns into another cell *j* after deformation.

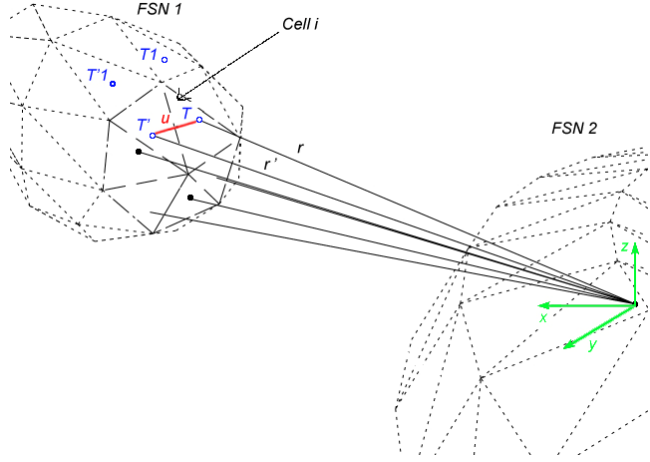

**Fig. 2.** Displacement of a *FD tag* from point *T* to point *T'* in an elastic lattice defined by the *FSN1* respect the *FSN2*

The point *T* with radius vector **r** before deformation – respect the ontology matching defined by another *FD tag* on *FSN2* – becomes point *T'* with radius vector **r'** after deformation, and **u** is the displacement vector of point *T* during the deformation process (see Fig. 2) as depict in (2).

$$(2) \quad \vec{r}' = \vec{r} + \vec{u}$$

The point *T1* becomes point *T'1* in cell *j'* after deformation. The radius vector connecting points *T1* and point *T'1* is d**r** = d*x'i* = d*xi* + d*ui*. The displacement of point *T1* is *u'*, thus (3).

$$(3) \quad \vec{u}' = \vec{u} + d\vec{u} \quad ; \quad du_i = u_i' - u_i$$

$$(4) \quad du_i = \frac{\partial u_i}{\partial x_i} dx_i \quad ;$$

We can express with (4) the Taylor expansion at point *T* taking the first order term only. Under the small deformation assumption, this reaches a very high accuracy.

The elastodynamic equations of the *FSN*, as extension to the classical elastodynamics formulation [14], can be deduced by considering the inertia effect respect the time exposition (*E*) depicted as partial derivative as in (5). Where *ρ* is the mass density of the *FD tags* and *e* is the time exposition of the single *FD tag*. Coherently the adaptive ontology matching, formulated before, and the elastodynamics ontology matching (5) are merged to obtain what we can call the *elasto-adaptative-dynamics methodology* of the *FSN*.

$$(5) \quad \frac{\partial \sigma_{ij}}{\partial x_i} + f_i = \rho \frac{\partial^2 u_i}{\partial e^2}$$

Omitting the body forces inside the *FSN* lattice and considering the equations (5) – with the deformation history [4 - 7] – we can express the generalized Hooke's as in (6), where *λ* represents the modulated displacement.

$$(6) \quad \sigma_{ij} = 2\mu\varepsilon_{ij} + \lambda\varepsilon_{kk}\partial_{ij}$$

**PRACTICAL USE**

The most practically useful part of this work, so far, is for the recommendation problem a so-called determinantal point process. But you can imagine another scenario where, for instance, you have a long document, and you want to take the five sentences that best summarize it. Or if you have a book and you want to generate a summary, then you need to take some pieces of the book and put them together. You want these frames to be representative of the whole book, while at the same time, they're not always the same thing.

**EXPERIMENTAL OBSERVATION**

For the experimental observations it was considered the data collection acquired analyzing the hashtags of Twitter [15]. Hashtags are an example of folksonomies for social networks; they are used to identify groups and topics (the short messages called Tweet in Twitter). Hashtags are neither registered nor controlled by any one user or group of users; they are used to identify groups and topics to identify short messages on microblogging social networking services such as: Twitter, Google+, YouTube, Instagram, Pintrest…

On December 2014 Twitter had surpassed 255 million monthly active users generating over 500 millions of tweets daily (source: Twitter).

For the Experimental Observation it was considered only 100K tweets connected to "world news" topics.

The *Folksodriven notation* was used for the content analysis on dynamic ontology matching for the Twitter #hashtags correlated to a topic *T* (for the evaluation on five different topics chosen). The system performance was measured in terms of efficiency of the analysis and matching process.

To measure the efficiency of the *Elastic Adaptive Ontology Matching* a stress test was done performing the two most expensive tasks occurring at run-time:

- the time needed by the algorithm to analyze a new set of 100K #hashtags from Twitter to be deployed and to generate the *Folksodriven tags* (*FD*),
- the time needed to automatically generate matching for the *Folksodriven Structure Network* (*FSN*)

Fig. 3 shows the distribution of response time obtained by issuing 500 Twitter #hashtags (*D*) for a single *Topic of interest* (*T*).

**RESULT VERIFICATION**

In order to consider the effectiveness of the approach presented here, six recent references, as potential benchmarks, are now tabulated in Table 1 and 2 to compare the results and consider the pros and cons of the proposed approach.

The potential benchmarks were chosen by the 2015 iteration of the SemEval-2015 Task 10 shared task on Sentiment Analysis in Twitter [15] that can be compared with the FSN algorithm proposed in this chapter. Three benchmarks were chosen by subtasks that ask to predict the sentiment towards a topic in a single tweet and the degree of prior polarity of a phrase (Subtask C of the SemEval-2015 Task 10).

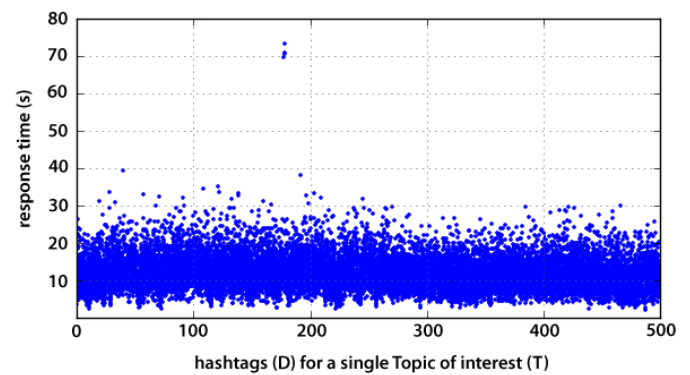

**Fig. 3.** Distribution of response time obtained by issuing 500 *#hashtags* (*D*) for a single *Topic of interest* (*T*)

|  | avgDiff | avgLevelDiff |
| --- | --- | --- |
| KLUEless | 0.202 | 0.810 |
| **FSN** | **0.204** | **0.824** |
| Whu-Nlp | 0.210 | 0.869 |
| TwitterHawk | 0.214 | 0.978 |

**Table. 1.** The proposed approach results verification w.r.t. the potential benchmarks according to the Subtask C of the SemEval-2015: Topic-Level Polarity. The systems are ordered by the official 2015 score.

|  | Kendall's τ coefficient | Spearman's ρ coefficient |
| --- | --- | --- |
| ONESC-ID | 0.6251 | 0.8172 |
| Ilsislif | 0.6211 | 0.8202 |
| **FSN** | **0.6109** | **0.8356** |
| ECNU | 0.5907 | 0.7861 |

**Table. 2.** The proposed approach results verification w.r.t. the potential benchmarks according to the Subtask E of the SemEval-2015: Degree of Prior Polarity. The systems are ordered by their Kendall's τ score, which was the official score.

Now, by considering the whole of factors in the cases, in a careful manner, it is clear to note that the proposed approach FSN is well behaved on Topic-Level Polarity (Subtask C of the SemEval-2015 Task 10), see Table 1. This proved to be a hard subtask: given a message and a topic, decide whether the message expresses a positive, a negative, or a neutral sentiment towards the topic. [16 - 19]

Moreover, on Table 2 is depicted the proposed approach FSN according the degree of prior polarity of a phrase (Subtask E of the SemEval-2015 Task 10) - determining Strength of Association of Twitter Terms with Positive Sentiment. Given a word/phrase, propose a score between 0 (lowest) and 1 (highest) that is indicative of the strength of association of that word/phrase with positive sentiment. If a word/phrase is more positive than another one, it should be assigned a relatively higher score.

## CONCLUSION

The present research attempts to address an efficient approach in the area of Computational Intelligence for Sentiment Analysis and Ontology Engineering. The main purpose of this chapter is the development of a constitutive model emphasizing the use of ontology-driven processing to achieve a better understanding of the contextual role of concepts (Sentic computing)..

In the approach proposed here, firstly, a new concept structure called "Folksodriven" to represent folksonomies is accurately identified. To face the problem of an adaptive ontology matching, a *Folksodriven Structure Network* (*FSN*) is built to identify the relations among the *Folksodriven tags*.

The present algorithm finds the deformation on *FSN* based on folksonomy tags chosen by different user on web site resources; this is a topic which has not been well studied so far.

The outlines have utilized to present a low-complexity approach, as long as the investigated results guarantee that the proposed approach is able to estimate how the properties of the *FSN* depend mainly on the nature, distribution, size and the quality of the reinforcing Folksodriven tags (*FD tags*).

Through a series of experiments, the ability of the approach is shown to be able to estimate the relations among the *FD tags* for adaptive folksonomies classifications in an evolving environment used by Intelligent Systems.

FD tags are used to categorize salient data points so they can be fed to a machine-learning system and "featurizing" the data. The discussion in the paper shows that the nonlinear elastic constitutive equation possesses some leaning for the investigation due to lack of plastic constitutive equation at present.

**Massimiliano Dal Mas** is an engineer working on webservices, trafficking and online advertising and is interested in knowledge engineering. In the last years he had to play a critical role at Digital Advertising business, cultivating relationships with key publisher partners with experience managing a team. Been responsible for all day to day operations with partners and consult on the best ways to monetize their properties. His interests include: user interfaces and visualization for information retrieval, automated Web interface evaluation and text analysis, empirical computational linguistics, text data mining, knowledge engineering and artificial intelligence. He received BA, MS degrees in Computer Science Engineering from the Politecnico di Milano, Italy. He won the thirteenth edition 2008 of the CEI Award for the best degree thesis with a dissertation on "Semantic technologies for industrial purposes" (Supervisor Prof. M. Colombetti). In 2012, he received the best paper award at the IEEE Computer Society Conference on Evolving and Adaptive Intelligent System (EAIS 2012) at Carlos III University of Madrid, Madrid, Spain. In 2013, he received the best paper award at the ACM Conference on Web Intelligence, Mining and Semantics (WIMS 2013) at Universidad Autónoma de Madrid, Madrid, Spain. His paper at W3C Workshop on Publishing using CSS3 & HTML5 has been recommended as position paper.